\documentclass[twocolumn,showpacs,aps,prl]{revtex4}
\usepackage{graphicx}  
\usepackage{dcolumn}   
\usepackage{bm}        
\usepackage{amssymb}   

\usepackage{epsfig,amsmath,graphicx,bbold}

\usepackage{mathtools}
\DeclarePairedDelimiter\abs{\lvert}{\rvert}
\DeclarePairedDelimiter\norm{\lVert}{\rVert}
\makeatletter
\let\oldabs\abs
\def\abs{\@ifstar{\oldabs}{\oldabs*}}

\let\oldnorm\norm
\def\norm{\@ifstar{\oldnorm}{\oldnorm*}}

\newcommand{\lt}{\left(}
\newcommand{\rt}{\right)}

\newcommand{\lqu}{\left[}
\newcommand{\rqu}{\right]}

\newcommand{\be}{\begin{equation}}
\newcommand{\ee}{\end{equation}}
\newcommand{\ba}{\begin{eqnarray}}
\newcommand{\ea}{\end{eqnarray}}
\newcommand{\fr}{\frac}
\newcommand{\nn}{\nonumber}
\newcommand{\bay}{\begin{array}}
\newcommand{\eay}{\end{array}}

\begin{document}

\title{Reconfigurable Optical Quantum Networks \\Using Multimode Quantum Frequency Combs and Pulse Shaping}

\author{Y. Cai$^{1,2}$, J. Roslund$^1$, G. Ferrini$^{1,3}$, F. Arzani$^1$, X. Xu$^2$, C. Fabre$^1$ and N. Treps$^1$}

\affiliation{$^1$ Laboratoire Kastler Brossel, UPMC-Sorbonne Universités, CNRS, ENS-PSL Research University, College de France; CC74, 4 Place Jussieu, 75252 Paris, France}

\affiliation{$^2$ State Key Laboratory of Precision Spectroscopy and Department of Physics, East China Normal University, 200062 Shanghai, China}

\affiliation{$^3$ Laboratoire Mat\'eriaux et Ph\'enom\`enes Quantiques, Sorbonne Paris Cit\'e, Univ. Paris Diderot, CNRS UMR 7162, 75013 Paris, France}

\date{\today}

\begin{abstract}
Multimode entanglement is quintessential for the design and fabrication of quantum networks, which play a central role in quantum information processing and quantum metrology. However, an experimental setup is generally constructed with a specific network configuration in mind and therefore exhibits reduced versatility and scalability.
The present work demonstrates an on-demand, reconfigurable quantum network simulator, using an intrinsically multimode quantum resource and a homodyne detection apparatus. Without altering either the initial squeezing source or experimental architecture, we realize the construction of thirteen cluster states of various size and connectivity as well as the implementation of a secret sharing protocol. In particular, this simulator enables the interrogation of quantum correlations and fluctuations for a Gaussian quantum network. This initiates a new avenue for implementing on-demand quantum information processing by only adapting the measurement process and not the experimental layout.
\end{abstract}


\maketitle

 
Inseparability, i.e. the impossibility of treating as separable entities physical systems that have been generated in an entangled, non-factorable quantum state, even though the systems are no longer coupled to each other by a physical interaction, is one of the most puzzling properties of the quantum world \cite{{Einstein1935},{Horodechi2009}}. 
The consequences of this quantum property have been harnessed in a range of applications, including quantum teleportation \cite{{Bouwmeester1997}, {Furusawa1998}} and quantum computation \cite{{DiVincenzo1995}, {Lloyd1999}}. 
In order to compete with classical computers, quantum computers need to employ a large number of quantum systems that are created in appropriately designed entangled states,  on which quantum processing operations can be performed before it is subject to decoherence. 
This multipartite quantum system is often termed a ``quantum network", and the individual quantum-correlated systems comprise the network ``nodes". The generation and use of large quantum networks raise numerous experimental and theoretical issues that are the subject of intense research. For instance, significant effort has recently been directed toward defining specialized metrics that assess the presence of multipartite entanglement \cite{{Amico2008}, {Huber2013}, {Levi2013}, {Sperling2013}} as well as characterize the ``quality" of such a quantum resource in view of quantum computing applications. This issue is still the subject of debate throughout the community.

The majority of hitherto studied systems have employed qubits (i.e. material-based two-level systems, such as ions or quantum dots) as the nodes of a quantum network. In this case, the parties comprising the multipartite quantum network are well-defined physical objects, and multipartite entanglement amongst nodes appears as a many-body property where each party is physically separated from the other and can be measured independently \cite{Kimble2008}. However, another variety of quantum multipartite system has recently emerged: the \textit{optical multimode systems},  which is the focus of the present work.

We consider the electric field quantum operator $\hat{E}^{(+)}(\bf{r},t)$ (a scalar field is assumed for simplicity), which is written in a more general form as: 
\begin{equation}
\hat{E}^{(+)}({\bf r},t)=\sum_i \hat{a}_i f_i ({\bf r},t),
\end{equation}
where $f_i ({\bf r},t)$ constitute a basis of optical modes (i.e. orthonormal solutions of Maxwell's equations with specific boundary conditions), $\hat{a}_i $ are photon annihilation operators in the mode of spatio-temporal shape $f_i ({\bf r},t)$, and an overall multiplicative factor has been ignored for simplicity. In this case, the entire quantum network is placed in an entangled, multimode quantum state. These states may include Fock or squeezed states but also various entangled states, such as two-mode Einstein-Podolski-Rosen (EPR) states, three-mode GHZ states, etc.

Compared to ``material-based quantum networks," photonic networks exhibit unique properties that include a relative insensitivity to decoherence but also an ability to arbitrarily change the mode basis. Toward this end, the field $\hat{E}^{(+)}$ may be rewritten as:
\begin{equation}
\hat{E}^{(+)}({\bf r},t)=\sum_j \hat{b}_j g_j ({\bf r},t),
\end{equation}
in which $\{g_j ({\bf r},t)\}$ represents another mode basis while $\hat{b}_j $ are the associated photon annihilation operators in the mode $g_j$. A transformation from the original modal basis and annihilation operators to another is accomplished by means of a unitary transformation:
\begin{equation}
\vec{g}=U^{\dagger} \, \vec{f} \quad ; \quad \vec{b}=U \, \vec{a},
\end{equation}
where $U$ is a unitary transformation acting on the vector space of modes, and the vectors $\vec{g}$, $\vec{f}$, $\vec{b}$, $\vec{a}$ have respective components $f_i$, $g_j$, $\hat{b}_j$, $\hat{a}_i$. The potential for examining a given quantum state in an arbitrary modal basis is unique to multimode quantum optics and does not exist for material qubits. Importantly, it is possible to experimentally access the properties of a given mode (e.g. $f_i$) using balanced homodyne detection in which a local oscillator is temporally and spatially sculpted in the same mode \cite{Polycarpou}. Such a measurement also has the potential to arbitrarily reconfigure the projection operator that acts on the multimode optical state of interest \cite{{Pysher2011}, {Armstrong2012}, {Chen2014}} in a spirit closely related to measurement based quantum computing \cite{{Lloyd1999}, {Mennicucci2006}, {Loock2007}}.

The properties of the quantum networks that are considered in this work are therefore not governed by the initial quantum state but rather by the measurement process. In particular, we demonstrate that multipartite entanglement is not only an intrinsic property of the source but also the result of a complex interplay amongst the source, act of measurement, and possibly post-processing that acts on the measurement results \cite{{Armstrong2012}, {Ferrini2013}, {Cai2015}}. This new avenue paves the way for configurable, adaptive, and scalable quantum information processing whose possibilities remain largely unexplored, both theoretically and experimentally.

While a multitude of experiments have demonstrated the construction of quantum networks, the experimental architecture typically realizes one specific structure and is not reconfigurable \cite{{Yokayama2013}, {Ukai2011b}, {Su2013}}. Hence, a general study on the diversity of networks that are attainable from a single fixed resource has not been performed. The present work tailors the local oscillator of multimode homodyne measurements, which enables accessing a multiplicity of networks without any modification of the experimental arrangement. As a result, a direct study is accomplished of the scalability and versatility of the networks that may be forged from a single resource.

After explaining how measurement based quantum networks can be implemented, we introduce the experimental platform, which is based upon parametrically-generated ultrafast frequency combs whose temporal/spectral structure is exploited to carry multimode quantum information \cite{Roslund2014}. The use of ultrafast pulse shaping combined with homodyne-based projective measurements allows the on-demand construction of various quantum networks. As a practical illustration, we focus on the generation of various cluster state structures that are fabricated from the same light resource. Subsequently, a multipartite quantum secret sharing proposal is implemented by making use of one of the generated cluster states.

\section{Results}

\begin{figure*}[htbp]
\centering
\includegraphics[width=135mm]{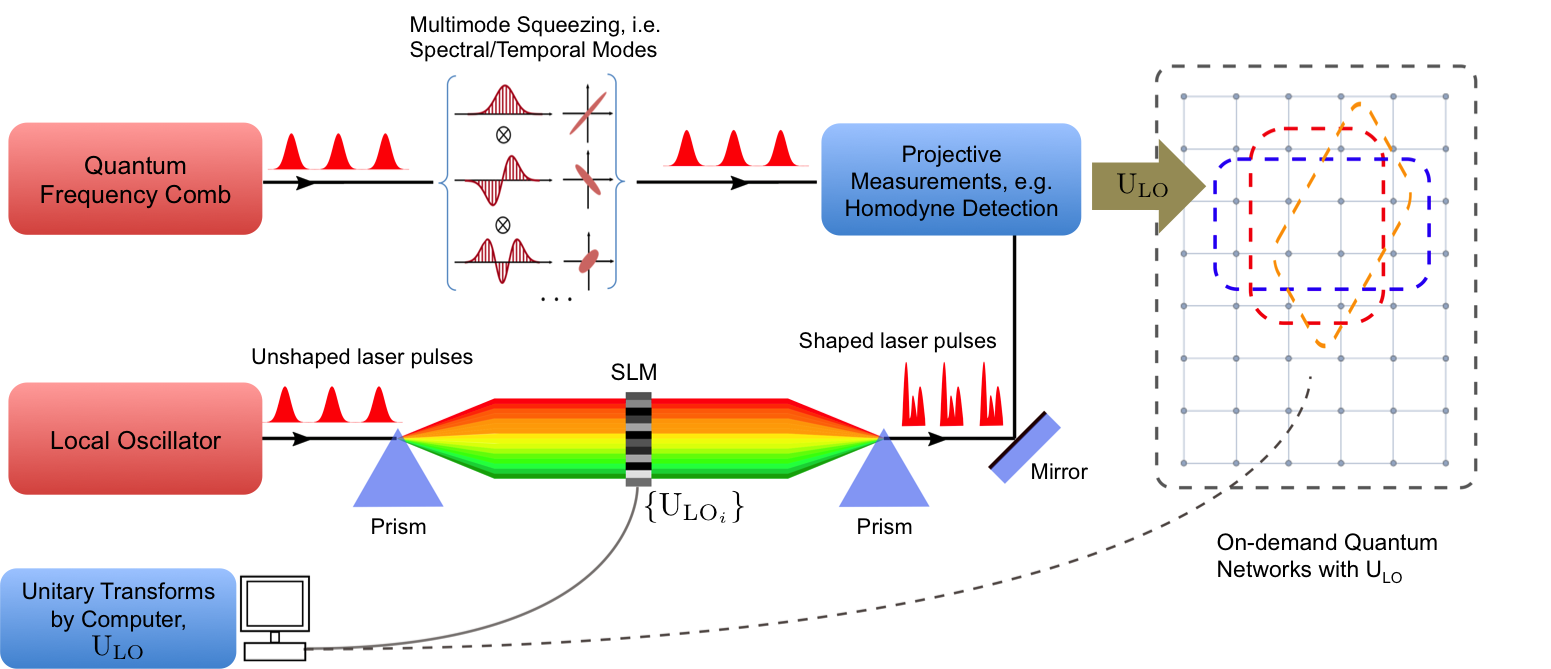}
\caption{\textbf{Experimental setup for the quantum network simulator.} The simulator is based upon a quantum frequency comb \cite{Pinel2012} and homodyne detection with a customized local oscillator. The quantum frequency comb is a multimode squeezed state, in which each squeezed mode possesses a specific spectral structure. Consequently, quantum correlations exist in the frequency-band basis \cite{Roslund2014}. On-demand quantum-network correlations within the frequency comb are revealed via projective measurements, which consists of homodyne detection in a suitable basis. The local oscillator (LO) is sculpted into the appropriate pulse shape by a computer-programmed spatial light modulator (SLM). The subsequent measurement of the quantum state with this shaped LO realizes the desired network unitary transformion $U_\textrm{LO}$. 
}
\label{figure1}
\end{figure*}

\textbf{Measurement based all-optical quantum-networks.} The Bloch Messiah decomposition \cite{Braunstein2005} states that any pure multimode Gaussian quantum state of light can be reduced to a set of uncorrelated squeezed vacuum states in an appropriately chosen mode basis of annihilation operators $\vec{a}^{\, \textrm{psqz}}$ (the array of modes are conventionally taken to all be squeezed in the $p$-quadrature of the field). This implies that the modes of any Gaussian quantum network may be constructed from a set of squeezed modes by implementing a proper change of mode basis. In practice, a network of interest may be fashioned by applying  a unitary transformation $U_{\textrm{net}}$ to a set of independently squeezed modes \cite{Loock2007}, which allows for the annihilation operators $\vec{b}^{\, \textrm{net}}$ of the network to be described as
\begin{equation}\label{qnetwork}
\vec{b}^{\, \textrm{net}} = U_{\textrm{net}} \, \vec{a}^{\, \textrm{psqz}}.
\end{equation}
The unitary transformation $U_{\textrm{net}}$ is conventionally implemented by means of a suitable arrangement of linear optical elements, including beamsplitters and phase shifters, and several pioneering experiments have demonstrated this approach \cite{{Su2013}, {Su2007}, {Yukawa2008}}.  As $U_{\text{net}}$ mathematically corresponds to a general basis rotation, an alternative, but equivalent, manner in which to reveal the optical network is to measure the multimode beam in the appropriate basis. Such a basis change can be implemented with a mode-selective detection system, which is the novel approach that is considered in this work.

Considering that any basis change is at hand, the realization of an arbitrary Gaussian quantum network may start from any highly multimode non-classical state. In the present situation, the parametric down conversion of an optical frequency comb generates full multimode entanglement in the frequency basis \cite{Gerke2015}. The spectral domain in which the downconversion occurs is described with a set of ``frequency-pixel modes" $\vec{h}^{\, \textrm{pix}}$  (in practice, they do not correspond to single frequency components but instead to a given frequency band matching the spectral resolution of the detection system)
with corresponding annihilation operators $\vec{a}^{\, \textrm{pix}}$. These optical modes constitute an approximate basis on which the squeezed modes $\vec{f}^{\, \textrm{psqz}}$ can be decomposed. The set of annihilation operators corresponding to the squeezed modes may then be written as $\vec{a}^{\, \textrm{psqz}} = U_{\textrm{sqz}}\vec{a}^{\, \textrm{pix}}$, where $U_{\textrm{sqz}}$ is the corresponding unitary transformation whose phase degrees of freedom are chosen such that $\vec{a}^{\,\textrm{psqz}}$ are squeezed along the $p$ quadrature. After applying the unitary transformation $U_{\textrm{net}}$ corresponding to the network of interest, the transformation becomes
\begin{equation}
\vec{b}^{\, \textrm{net}} = U_\textrm{net} \, U_{\textrm{sqz}} \, \vec{a}^{\, \textrm{pix}}=U_\textrm{LO} \vec{a}^{\, \textrm{pix}}.
\label{network}
\end{equation}
Consequently, every quantum network possesses a unitary matrix $U_\textrm{LO}$ that allows it to be related to the frequency-pixel mode basis. As seen in Fig. \ref{figure1}, this transformation can be implemented by a series of homodyne measurements with the Local Oscillator (LO) in the appropriate spectral shape. The quality of the extracted network itself depends on the amount of squeezing in each mode, and thus also on an optimized choice of the $U_{\textrm{net}}$ matrix \cite{Ferrini2015}.


\textbf{The Quantum Resource. }The multimode quantum resource is formed from the parametric downconversion of an ultrafast pulse train. A 76 MHz pulse train delivering $\sim$120 fs pulses centered at 795 nm is frequency doubled, which serves to pump a $\chi^{(2)}$ non-linear crystal in a low finesse cavity. This pump source is composed of about $\sim 10^5$ single frequencies, each of which can be the potential source of $\sim 10^5$ different pairs of down-converted photons \cite{Valc‡rcel2006} . The resultant downconverted source can be characterized by either directly assessing its entangled character in the frequency domain or by extracting the eigenmodes of the downconversion process \cite{{Gerke2015}, {Pinel2012}, {Roslund2014}}. Given the highly multimode character of the downconverted comb, the limits of the quantum resource are determined by the quality of the detection process \cite{Araujo2014}.

 \begin{figure*}[htbp]
\centering
\includegraphics[width=180mm]{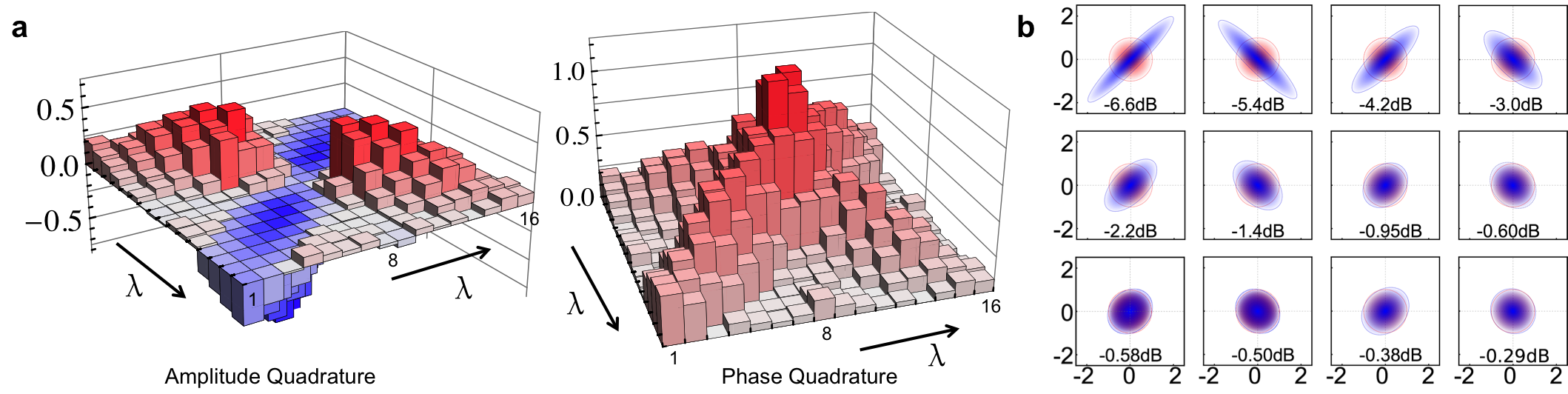}
\caption{\textbf{Multimode quantum resource.} (a) 16-partite covariance matrix in the frequency-pixel basis. This correlation matrix is obtained with balanced homodyne detection where the spectrum of the local oscillator is divided into 16 frequency bands of equivalent width.  The shot noise contribution has been subtracted from the diagonal for increased visibility. (b) The squeezing ellipses (blue) of the twelve leading squeezed eigenmodes. The circles (red) represent vacuum fluctuations for comparison. Twelve of the sixteen modes are squeezed. The measurement results in panels (a) and (b) have been corrected for electrical dark noise and a 15$\%$ optical loss in the measurement processes. }
\label{figure2}
\end{figure*}

Detection is performed with pulse shaped homodyne detection (see Fig. \ref{figure1}). In order to reach a highly multimode regime, a high resolution pulse shaper is used along with high quantum efficiency detectors (see Methods). The resolution of the pulse shaper is $\sim0.06$ nm/pixel in a 30 nm band centered at 795 nm. The LO field, which originates from the same source laser, undergoes both amplitude and phase spectral shaping with this device, and the resulting shape defines the detection mode of the homodyne setup.

In order to characterize the initial quantum resource, the LO spectrum is first divided into 16 frequency bands of equal bandwidth ($\sim$ 0.8 nm). These bands correspond to the pixel modes of equation (\ref{network}). With the same general strategy as the one presented in \cite{Roslund2014} except for a direct computer acquisition of the noise data (see Methods), the accurate measurement of large covariance matrices is accomplished in a short time period (around 1s). 
The resultant amplitude and phase covariance matrices are shown in Fig.~\ref{figure2}a. By applying a Bloch-Messiah decomposition to these matrices  \cite{Braunstein2005}, 16 eigenvectors and eigenvalues are extracted, which correspond to the orthogonal squeezed spectral modes and their respective squeezing values (see Fig.~\ref{figure2}b). These modes comprise the input basis of our quantum network, which consists of 12 significantly squeezed modes with squeezing values ranging from -0.3 dB to -6.6 dB. The squeezing values presented in Fig.~\ref{figure2}b are corrected for the homodyne detection visibility and detection losses (15$\%$ in total, including the visibility) since they represent the initial quantum resource. Henceforth, only the dark noise contribution to the data is removed for the networks presented in the remainder of this work, and no correction is applied for detection losses.

To summarize, the process of parametric downconversion provides the link between the 16-mode operators in the measurement basis $\vec{a}^{\, \textrm{pix}} = (\hat{a}_{1}^{\textrm{pix}}, ...,\hat{a}_{16}^{\textrm{pix}} )$ and those in the squeezed basis $\vec{a}^{\, \textrm{psqz}} = (\hat{a}_{1}^{\textrm{sqz}}, ...,\hat{a}_{16}^{\textrm{sqz}} )$. This link is the experimentally measured unitary transformation $U_{\textrm{sqz}}$, which acts as $ \vec{a}^{\, \textrm{psqz}} = U_{\textrm{sqz}} \, \vec{a}^{\, \textrm{pix}}$. In order to reveal any quantum network, the local oscillator is shaped according to equation (\ref{network}). The high resolution of the pulse shaper allows for a fine reconstruction of the network, at the expense of detecting only one mode at a time. This system is thus a {\it quantum network simulator}, as it allows for accessing any of the modes or witness inequalities (see next section) that characterize a given quantum network, but with the restriction that they cannot be revealed simultaneously.

\begin{figure*}[htbp]
\centering
\includegraphics[width=165mm]{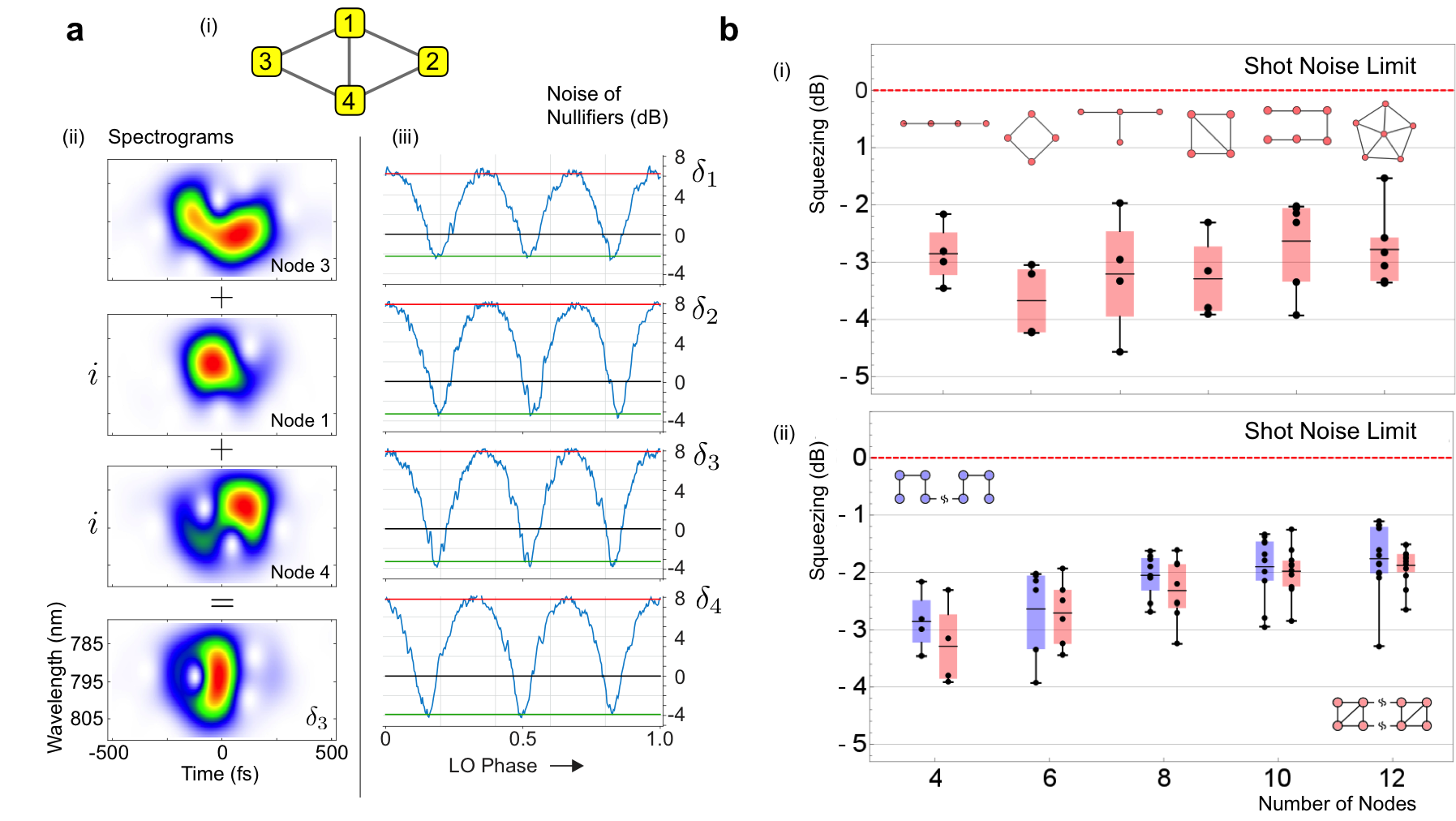}
\caption{\textbf{Simulation of cluster states.} (a) Witness measurement. (i): graph of the diagonal-square four-node cluster state as defined by the adjacency matrix $V$. (ii): A sample of the spectrograms that correspond to this cluster state. The leading three spectrograms depict the pulse shapes corresponding to the optical nodes indicated in the corner of each image, and the final spectrogram represents the nullifier $\delta_3$ for node 3 as defined by equation (\ref{eq:nullifiers}), which is formed as the spectral superposition $\hat{a}_{3}^{\textrm{null}} = \hat{a}_{3} + i \, \hat{a}_{1} + i \, \hat{a}_{4}$. (iii): In order to measure the nullifier variances associated with this cluster state, the pulse shaper sculpts the LO in the spectral form associated with each of the four nullifiers, and the resulting 
four variance curves as a function of the global LO phase are shown. Each of the four nullifiers exhibits noise statistics below the shot noise limit (black lines). 
(b) Versatility and scalability. Nullifier squeezing values of various cluster states possessing between 4 and 12 nodes are presented as box plots. The black points are the individual nullifier variances, the pink rectangles depict the first and third quartiles of the data, the black line contained in the rectangle is the nullifier mean, and the black whiskers indicate the upper and lower extrema of the nullifier collection. The red dashed lines in (i) and (ii) represent the shot noise limit. 
All of the nullifier variances are below the shot noise limit, which implies successful generation of the targeted cluster states. In (ii), the variances of the n-mode linear (left, blue) and diagonal square (right, pink) cluster states are compared. The noise variances in panels A and B are only corrected for electrical dark noise.}\label{figcluster} \end{figure*}

\textbf{Cluster state simulations. }With an eye towards applications in measurement based quantum computing \cite{{Lloyd1999}, {Mennicucci2006}, {Loock2007}}, we first simulate a series of different cluster states. Cluster states are highly entangled multimode Gaussian states for which specific quadrature combinations, called \emph{nullifiers}, are defined by 
\begin{equation} 
\label{eq:nullifiers}
\vec{\delta} = \vec{p}^{\, C} - V \cdot \vec{x}^{\, C},
\end{equation}
and should satisfy the relation $\Delta^2 \vec{\delta} \to 0$ in the limit that input squeezing levels are infinite \cite{{Mennicucci2006}, {Loock2007}}. In this formulation, $\vec{x}^{\,C}$ and $\vec{p}^{\,C}$ are, respectively, the amplitude and phase quadratures of the cluster nodes $\vec {a}^{\,C}= \vec{x}^{\,C} + i \, \vec{p}^{\,C}$, and $V$ is the adjacency matrix of a graph that defines the connectivity of the cluster state. 

A unitary transformation $U_{C}$ may be used to represent each cluster node as a complex superposition of the uncorrelated squeezed states embedded within the comb output. The individual nullifier relations as defined by equation (\ref{eq:nullifiers}) also correspond to specific superpositions of the squeezers. Consequently, a particular spectral mode may be associated with each of these nullifiers. As an example, the pulse shapes that characterize each node of a diagonal-square four-node cluster state are shown in Fig. \ref{figcluster}a. The optical mode corresponding to the third nullifier $\delta_3$ is constructed by shaping the LO into a form that corresponds to the summation of the amplitude quadrature of cluster node three with the phase quadratures of cluster nodes one and four (as specified by equation~\ref{eq:nullifiers}), i.e. $\hat{a}_{3}^{\textrm{null}} = \hat{a}_{3} + i \, \hat{a}_{1} + i \, \hat{a}_{4}$. This shaped LO pulse form is projected onto the multimode entangled state by homodyne detection, which allows for measuring the nullifier variance of the associated cluster node. The nullifier variances for the other modes are obtained in a similar fashion. As seen in Fig. \ref{figcluster}a, all of the nullifiers variances possess squeezing values between -2 and -4 dB, which indicates successful generation of this cluster state by the simulator.

This scheme was also exploited to fabricate additional cluster states with nodes that range in number from four to twelve. In Fig.~\ref{figcluster}b (i) (top), the nullifiers corresponding to a number of 4- and 6-node cluster networks are represented along with the corresponding connectivity structure. These variances are once again measured by a suitable programming of the pulse shaper as prescribed by equation~\ref{eq:nullifiers}, in which the adjacency matrix $V$ for each cluster is given by the geometrical figure above the corresponding nullifier. Additionally, the scalability with respect to cluster dimensionality is analyzed in Fig.~\ref{figcluster}b (ii) (bottom), where linear and diagonally-connected square clusters are constructed from a number of nodes that ranges from four to twelve. Both of these structures possess a set of nullifiers that lie below the shot noise limit for all considered dimensionalities, which indicates successful creation of the various networks. 

Importantly, the unitary transformation $U_C$ leading to a given cluster state is not unique. For the situation in which each of the input squeezers possesses the same degree of squeezing, a basis rotation on these modes prior to the $U_C$ transformation would not change the obtained network \cite{Ferrini2015}. However, in the case of disparate input squeezing levels, the measured nullifier variances depend upon the specific choice of the unitarity transformation. 
The present work optimizes the choice of the network matrix $U_{\, C}$ among all of the possible basis rotations that yield a given cluster structure \cite{Ferrini2015}. This is accomplished with an offline optimization that minimizes the mean of the cluster variances for a specific structure given the experimental input squeezers of Fig.~\ref{figure2}. As a result, the mean of the nullifier variances is approximately equal across the examined cluster series as seen in Fig.~\ref{figcluster}b, which indicates that the finite resources available have been optimally allocated (see also the Method section).

Among the variations that persist following optimization, it is observed that cluster states with a higher connectivity exhibit a lower mean nullifier variance for a fixed number of modes. For example, the mean nullifier variance of the four-node square graph state shown in Fig.~\ref{figcluster}b (i) is seen to be lower than those of the four-mode linear and T-shape graphs. Likewise, the six-node star cluster state has a lower mean variance than that of the six-node linear structure. This observation is also readily apparent in Fig.~\ref{figcluster}b (ii) in which the diagonally-connected square cluster states consistently have lower variances than linear ones for a given number of modes. Intuitively, one can understand that a highly connected state makes better use of resources. This corresponds to the fact that the requisite degree of entanglement contained within a state depends upon the quantum information task that is to be performed \cite{Gross2009}.

\textbf{Quantum secret sharing simulations. }Quantum secret sharing consists of sharing information (either quantum or classical) between several \emph{players} through the use of entangled quantum states. The information is first transferred to a multipartite entangled state. Each player is then given a piece of the total entangled state, and the original information can only be retrieved through a collaboration of subsets of the players. The quantum correlations increase both the protocol security as well as its retrieval fidelity as compared to what is attainable with only classical resources~\cite{{Hillery1999}, {Lance2004}, {Markham2011}}.

\begin{figure*}[htbp]
\centering
\includegraphics[width=130mm]{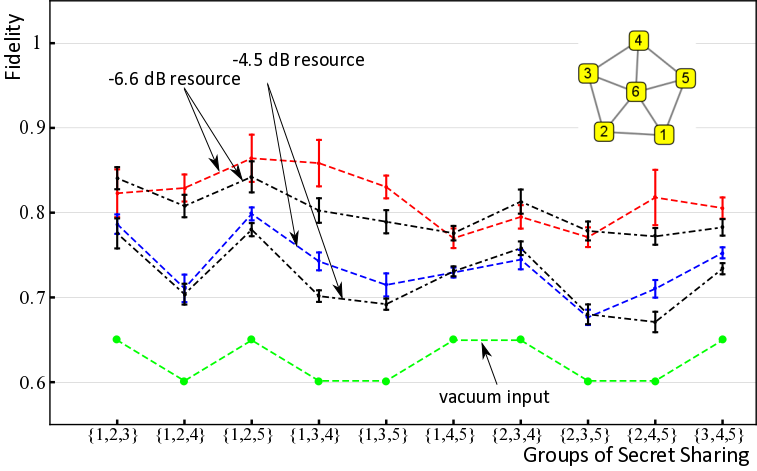}
\caption{\textbf{Experimental fidelities for the simulation of quantum secret sharing.} The graph state that is used in the secret sharing protocol is shown as an inset in the upper right portion of the figure. Nodes 1 to 5 constitute the players and node 6 is the dealer. The horizontal axis of the plot shows all 10 of the possible combinations for a three member access party. The red and blue dashed lines are the measured fidelities of the reconstructed secret given a -6.6 dB and -4.5 dB squeezing resource, respectively. The black dashed-point curves are inferred from the individual squeezing of the eigenmodes through the use of equation~\ref{secretsharing}. The green dashed curve corresponds to the fidelity with ordinary vacuum replacing the squeezed resource. All of these fidelities are directly measured and only contain a correction for the electrical dark noise. }
\label{figsecret}
\end{figure*}

Here we emulate a five-partite secret sharing protocol, which uses a six mode quantum network with the graph structure shown in Fig.~\ref{figsecret}. Nodes on the edge of the pentagon (labeled 1 to 5) represent the players, and the central node (6) encodes the secret prior to its coupling to the conglomerate state. Hence, this central information carrying node is termed the \emph{dealer}.

In the present case, the nodes corresponding to the players and the dealer are associated with the annihilation operators $\hat{a}^\textrm{net}_{i}$, which, in turn, are  constructed as a combination of the leading six squeezed eigenmodes of the comb. This transformation is obtained with the same matrix $U_\textrm{6se}$ that is employed to build the rightmost cluster state in Fig.~\ref{figcluster}b (i). This choice was inspired by \cite{LoockMarkham}, and the total transformation is written as:

\be
\vec{a}^\textrm{net}=\lt\begin{array}{c}
\hat{a}^\textrm{net}_1\\
\hat{a}^\textrm{net}_2\\\hat{a}^\textrm{net}_3\\\hat{a}^\textrm{net}_4\\\hat{a}^\textrm{net}_5\\\hat{a}^\textrm{net}_\textrm{6, dealer}\\
\end{array}\rt
=U_\textrm{6se}\cdot \lt\begin{array}{c}
\hat{a}^\textrm{sqz}_1\\
\hat{a}^\textrm{sqz}_2\\\hat{a}^\textrm{sqz}_3\\\hat{a}^\textrm{sqz}_4\\\hat{a}^\textrm{sqz}_5\\\hat{a}_\textrm{s}\\
\end{array}\rt,
\label{secret}
\ee
where the operators $a^\mathrm{sqz} _i$ are the annihilation operators for the leading five squeezed eigenmodes of the quantum resource, as defined in previous sections. The sixth squeezed mode comprises the secret state, i.e. $\hat{a}_\textrm{s} = a^\mathrm{sqz}_{6}$.

Given this configuration, at least three players must collaborate to reconstruct the secret (see Methods for details). Any set of three players constitutes what is termed an \emph{access party}. As an example, we consider the access party of players one, two and three. In order to access and therefore reconstruct the $\hat{x}_s$ or $\hat{p}_s$ field quadrature of the secret state, the three players within this access party must each measure a specific quadrature of their local fields $\hat{a}^\textrm{net}_{i}$, and combine their independently obtained results with the dealer's $p$ quadrature measurement in the following access party operators:
\ba
\hat{x}_{123}&=&\sum_{i=1}^3 m_i \hat{x}^\textrm{net}_i+\sum_{j=1}^3 n_i \hat{p}^\textrm{net}_i+ C \hat{p}^\textrm{dealer}\nn \\
\hat{p}_{123}&=&\sum_{i=1}^3 p_i \hat{x}^\textrm{net}_i+\sum_{j=1}^3 q_i \hat{p}^\textrm{net}_i+ D \hat{p}^\textrm{dealer},
\label{secretsharing1}
\ea
where the coefficients $m_{i}, n_{i}, p_{i}$, $q_{i}$, $C$ and $D$ are real. The value of these coefficients, and thus the specific linear combination between the measurements, is dictated by the condition that the final result must contain only field quadratures of the secret as well as squeezed quadratures of the input resource. Importantly, any linear combination that results in the measurement of an anti-squeezed quadrature of the input resource must be avoided. These conditions ensure that in the limit of infinite squeezing, the statistics of the measurement precisely reflect those of the secret state. After rewriting the access party quadrature measurements under these conditions, one finds the following form for the access party operators:
\ba
\hat{x}_{123}&=&\hat{x}_s+\sum_{i=1}^5 a_i ^\mathrm{sqz} \hat{p}_i^\mathrm{sqz}\nn\\ 
\hat{p}_{123}&=&\hat{p}_s+\sum_{i=1}^5 b_i ^\mathrm{sqz} \hat{p}_i^\mathrm{sqz}.
\label{secretsharing}
\ea
Thus, the combined measurements of the access party and the dealer yield an estimation of the secret, whose retrieval fidelity directly depends upon the degree of input squeezing and the choice of the $U_{\textrm{6se}}$ matrix. More precisely, if the unitary $U_\textrm{6se}$ is completely general (i.e. not associated with the pentagonal cluster examined in the present case), it is not guaranteed that the access party quadrature combinations can be written in a form consisting of only squeezing quadratures of the resource state as in equation~(\ref{secretsharing}). For the situation in which such a form is indeed possible, the corresponding network may be utilized for secret sharing as in the present case. It also then becomes possible to demonstrate that no solution exists for groups of only two players, which implies that two players alone can not recover the secret by virtue of the fact that the contribution of the anti-squeezing quadratures can not be fully removed, which corrupts their individual measurements even in the limit of infinite squeezing (see Methods for details).

In a genuine secret sharing scenario, $\hat{p}^\textrm{dealer}$ is measured first, and the result is broadcast to the players via a classical channel, thus implementing the encoding of the secret state onto the players graph. In our case, the quadratures of the secret are reconstructed by shaping the LO to coincide with the linear combination of resource modes described in equation~(\ref{secretsharing1}).  In order to assess the quality of the secret sharing simulation, we measured the residual noise associated with $\hat{x}_{123}$ and $\hat{p}_{123}$ (see Methods for details). These noise variances are measured for two different multimode squeezing resources. For the first case, the quantum source is operated in a configuration that contains  -6.6 $\textrm{dB}$ (corrected for losses) of squeezing in the leading mode (this corresponds to the squeezers seen in Fig.~\ref{figure2}b.). In the second case, the overall squeezing is decreased by appropriately adjusting the pump power driving the parametric process, such that the leading squeezer exhibits a noise reduction of -4.5 $\textrm{dB}$ (corrected for losses) relative to the vacuum level. The distribution of noise variances for the squeezers is the same in both situations and follows the trend observed in Fig.~\ref{figure2}b. 
The retrieval fidelities for all 10 possible access party combinations are determined by measuring the noise variances prescribed by equation~(\ref{secretsharing1}) and are displayed in Fig.~\ref{figsecret}. For purposes of comparison, the same access party noise variances are also measured in the absence of squeezing (i.e. the resource state is a vacuum state), which are also shown in Fig.~\ref{figsecret}. As expected, the mean value for these retrieval fidelities is $\sim0.60$, which corresponds to the classical limit \cite{{grangier}, {Tyc}} (more details in Methods). With quantum resources, however, we observe fidelities higher than the classical limit, which increase with better squeezing.

The accuracy with which the pulse shaper sculpts the field combinations dictated by equation~(\ref{secretsharing1}) is also assessed by directly calculating the expected fidelities based upon the known input squeezing levels with the help of equation~(\ref{secretsharing}). These calculated fidelities are displayed as the black curves in Fig.~(\ref{figsecret}) for each of the two utilized multimode resources. The agreement between these calculated fidelities and the experimentally-measured ones is generally good. The origin of deviations between the two curves arises from the fact that spectrally-dependent losses encountered in the production and detection of the multimode state do not allow the amplitude and phase quadratures of the covariance matrix to be simultaneously diagonalized \cite{Roslund2014}. As a result, the spectral form of the eigenmodes for the two quadratures is slightly different, and this corrupts the perfect cancellation of the anti-squeezing contribution in equation~(\ref{figsecret}). This effect is more present with a higher level of squeezing, as the influence of losses becomes more significant. In principle, these deviations may be reduced by minimizing spectrally-dependent losses in the generation and detection of the quantum source. Nonetheless, the general agreement between the experimentally-measured and calculated variances confirms the utility of the apparatus at simulating arbitrary mode constructions. Despite the fact that the input secret can not be varied, as is usual in demonstrations of quantum secret sharing, this study allows for implementing secret sharing protocols consisting of a large number of modes while also exploring the influence of parameters such as loss and squeezing values.

\section{Discussion}
In summary, we have experimentally implemented a versatile and scalable detection scheme that allows for on-demand simulation of arbitrary quantum optical networks. This approach permits a direct interrogation of all of the relevant information that characterizes a multimode Gaussian state in a user-defined basis. Examples of such a synthesis include cluster state generation as well as a multipartite quantum secret sharing protocol that is built upon a six-node cluster network. 

Importantly, the creation of these cluster structures with our simulator does not necessitate any change in the optical architecture. Rather, the connectivity of the network structure is varied by simply modifying the basis in which the state is detected. Given that an arbitrary, multimode Gaussian transformation of a set of squeezers can be achieved with a unitary matrix, a set of identifiable pulse shapes may be associated with the transformation output. In this manner, it is possible to directly probe any Gaussian entanglement criteria. 
The fact that each of these structures is revealed by only adjusting the measurement basis indicates that these networks are all implicitly embedded within the multimode entangled resource. 

In other words, spectral-selectivity in the detection process, which may be accomplished with pulse shaping, or more generally with a variety of spectrally resolved homodyne detection processes, allows the synthesis of any linear combination of the individual squeezed modes present in the light source. Furthermore, the combination of this approach with multipixel homodyne detection is promising for multiple applications, including measurement-based quantum computing \cite{Ferrini2013} \cite{Ferrini2015}.  This is another evidence of the unique flexibility of the quantum system that we have chosen to study: femtosecond pulse trains as a basis for quantum networks.

\section{methods}

\label{sec:methods}

\textbf{Detection and data acquisition. } Light detection is achieved with balanced homodyne detection, which is performed with selected silicon photodiodes that exhibit $\sim$ 99$\%$ quantum efficiency and a bandwidth of $\sim 100$MHz. The homodyne fringe visibility is $\sim$ 93-95$\%$, and the total loss for the detection of squeezing is $\sim 15 \%$.  The photocurrent difference is amplified with a commercial amplifier (model Mini-Circuits ZFL-500LN) and then demodulated at 1 MHz. Each squeezing curve is measured following $\sim 1$~sec of data acquisition. Hence, an $n$-dimensional covariance matrix is fully measured in $n \cdot (n+1) / 2$~secs or $\sim 2$~mins for the 16-dimensional matrix shown in the present work. 


\textbf{Optimization of unitary cluster matrix. }For cluster states, one can demonstrate that if $U_\textrm{net}$ in equation~(\ref{qnetwork}) is a unitary matrix that leads to a cluster defined by its adjacency matrix \cite{Ukai2011b}, then the application of an arbitrary orthogonal matrix $\mathcal{O}$ to the unitary matrix (i.e. $U_\textrm{net}\mathcal{O}$) also leads to the same graph cluster state \cite{Ferrini2015}. Due to the non-uniform squeezing distribution of our multimode quantum resource (as seen in Fig.~\ref{figure2}), the measured nullifier variances are dependent upon the specific choice of the unitary transformation. In order to equally distribute the finite squeezing resources amongst the targeted cluster, an evolutionary algorithm is utilized to search for the matrix $\mathcal{O}$  that minimizes the mean nullifier variance based upon the measured covariance matrix.

\textbf{Unitary transform and fidelity of quantum secret sharing protocol. }For the secret sharing protocol presented in Fig. \ref{figsecret},  the corresponding six-node cluster matrix $U_\textrm{se}$ used in equation (\ref{secret}) has real part $X_\mathrm{se}$ \be \left(
\begin{array}{cccccc}
 .6234 & .0078 & -.1375 & -.1375 & .0078 & -.0591 \\
 .0078 & .6234& .0078 & -.1375 & -.1375 & -.0591 \\
 -.1375 & .0078 & .6234 & .0078 & -.1375 & -.0591\\
 -.1375 & -.1375& .0078 & .6233 & .0078 & -.0591 \\
 .0078 & -.1375 & -.1375 & .0078 & .6234 & -.0591 \\
 -.0591 & -.0591 & -.0591 & -.0591 & -.0591 & .4822 \\
\end{array}
\right),\nn\ee and the corresponding imaginary part, $Y_\mathrm{se}$, is \be \left(
\begin{array}{cccccc}
 -.0434 & .4268 & -.1887 & -.1887 & .4268 & .3641 \\
 .4268 & -.0434 & .4268 & -.1887 & -.1887 & .3641 \\
 -.1887 & .4268 & -.04342 & .4268& -.1887& .3641 \\
 -.1887& -.1887& .4268& -.0434 & .4268& .3641 \\
 .4268 & -.1887& -.1887& .4268& -.04342& .3641\\
 .3641& .3641 & .3641& .3641& .3641& -.2954\\
\end{array}
\right).\nn \ee Its action on the quadrature operator is represented by the symplectic matrix \be S_\mathrm{se}= \left( \begin{array}{cc} X_\mathrm{se} & -Y_\mathrm{se} \\ Y_\mathrm{se} & X_\mathrm{se} \end{array} \right). \ee The network quadrature operators are then obtained as \begin{eqnarray} \hat{x}^\mathrm{net} _i &= \sum \limits _{j=1} ^6 \left( X_{\mathrm{se},ij} \hat{x}^{\mathrm{sqz}}_j -  Y_{\mathrm{se},ij} \hat{p}^{\mathrm{sqz}}_j \right) \label{xnetcond} \\ \hat{p}^\mathrm{net} _i &= \sum \limits _{j=1} ^6  \left( Y_{\mathrm{se},ij} \hat{x}^{\mathrm{sqz}}_j + X_{\mathrm{se},ij} \hat{p}^{\mathrm{sqz}}_j \right) \label{pnetcond}, \end{eqnarray} which are actually a set of twelve equations expressing the local quadratures given to the players ($i = 1, ...,5$) and the dealer ($i = 6$). The secret is encoded in the sixth squeezed mode. To explain how the secret quadratures are measured by an access party, let us concentrate on a specific one, namely the one composed by players one, two and three as in the main text. Players are allowed to measure either the local position or momentum quadrature, or a rotated version of the two. They may then collaborate, linearly combining their outcomes. Moreover, the dealer measures $\hat{p}^{\mathrm{dealer}}$ and broadcasts the result to all the players. In practice, our experiment measured the local quadratures of each access party and the dealer's momentum quadrature at the same time by a suitable shaping of the local oscillator; nonetheless, we will detail the procedure to retrieve the secret quadrature in the scenario outlined above. Importantly, the result does not change.\\

Let us consider the access party of players one, two and three. Assume that the dealer measures $\hat{p}^\mathrm{dealer} = \hat{p}^\mathrm{net}_6$ getting the result $\mu$. As a consequence, the last terms of equations~(\ref{xnetcond}) and (\ref{pnetcond}) dictate a relation between the initially squeezed quadratures and the secret quadratures. We can use this new relation to rewrite one of the anti-squeezed quadratures, say $\hat{x}^\mathrm{sqz}_1$ in terms of $\mu$, the five remaining anti-squeezed quadratures $\hat{x}^\mathrm{sqz}_i$, and all six of the squeezed quadratures $\hat{p}^\mathrm{sqz}_i$. The first three components of both equations~(\ref{xnetcond}) and~(\ref{pnetcond}) are rewrritten as ($i=1,2,3$)\begin{eqnarray} \hat{x}^\mathrm{net} _i &= \sum \limits _{j=2} ^6   X'_{\mathrm{se},ij} \hat{x}^{\mathrm{sqz}}_j - \sum \limits _{j=1} ^6  Y'_{\mathrm{se},ij} \hat{p}^{\mathrm{sqz}}_j  + A\mu \\ \hat{p}^\mathrm{net} _i &= \sum \limits _{j=2} ^6   Y''_{\mathrm{se},ij} \hat{x}^{\mathrm{sqz}}_j + \sum \limits _{j=1} ^6  X''_{\mathrm{se},ij} \hat{p}^{\mathrm{sqz}}_j + B\mu,  \end{eqnarray} where $A$ and $B$ are real numbers. In order to reconstruct one of the secret quadratures, say $\hat{x}_s = \hat{x}^{\mathrm{sqz}}_6$, the players need to consider linear combinations of the local operators $\hat{x}^\mathrm{net}_i$ and $\hat{p}^\mathrm{net}_i$ of the form \be \begin{array}{rl} \hat{x}^{123} =&\sum \limits _{i = 1} ^3 m_i \hat{x}^\mathrm{net}_i + \sum \limits _{i = 1} ^3 n_i \hat{p}^\mathrm{net}_i \\ 

=& \sum \limits _{j=2} ^6 \sum \limits _{i = 1} ^3 \left( m_i X'_{\mathrm{se},ij} + n_i Y''_{\mathrm{se},ij}\right) \hat{x}^{\mathrm{sqz}}_j \\

 & + \sum \limits _{j=1} ^6 \sum \limits _{i = 1} ^3 \left( n_i X''_{\mathrm{se},ij} - m_i Y'_{\mathrm{se},ij}\right) \hat{p}^{\mathrm{sqz}}_j + C\mu.

\end{array}
\ee
 $C$ is a real number which depends on the coefficients $m_i$ and $n_i$. The goal of the players is to find coefficients $m_i$ and $n_i$ such that
 
 \be 
 \left\{
 \begin{array}{rl} 
 \sum \limits _{i = 1} ^3 \left( m_i X'_{\mathrm{se},ij} + n_i Y''_{\mathrm{se},ij}\right) =& 0 \textrm{    for    } j = 2,3,4,5 \\
\sum \limits _{i = 1} ^3 \left( m_i X'_{\mathrm{se},ij} + n_i Y''_{\mathrm{se},ij}\right) =& 1 \textrm{    for    } j = 6\\
\sum \limits _{i = 1} ^3 \left( n_i X''_{\mathrm{se},ij} - m_i Y'_{\mathrm{se},ij}\right) =& 0 \textrm{    for    } j = 6.
 \end{array} \right . \label{system}
 \ee
 As such, $\hat{x}^{123}$ will not contain the anti-squeezed quadratures, and the coefficient of the secret momentum quadrature $\hat{x}_s$ is one. If a solution of the linear system~(\ref{system}) exists, the access party has access to the measurement of \be \hat{x}^{123} = \hat{x}_s + \sum \limits_{i=1} ^5 a_i \hat{p}^{\mathrm{sqz}}_i + C\mu \ee where the $a_i$'s are fixed by the solution of~(\ref{system}). The real number $C\mu$ is known since $\mu$ is broadcasted by the dealer. Thus, with classical post-processing, the access party can measure \be \hat{x}^{123} = \hat{x}_s + \sum \limits_{i=1} ^5 a_i \hat{p}^{\mathrm{sqz}}_i. \ee A similar reasoning allows the access party to measure $\hat{p}^{123}$ as in the main text. We checked numerically that a solution exists for both $\hat{x}^{ijk}$ and $\hat{p}^{ijk}$ for every possible access party. Also, we verified that no solution exists when any pair of players is considered. Consequently, no less than three players can avoid the anti-squeezed quadratures, which spoils a retrieval of the secret quadrature.\\

\begin{figure}[htbp]
\centering
\includegraphics[width=85mm]{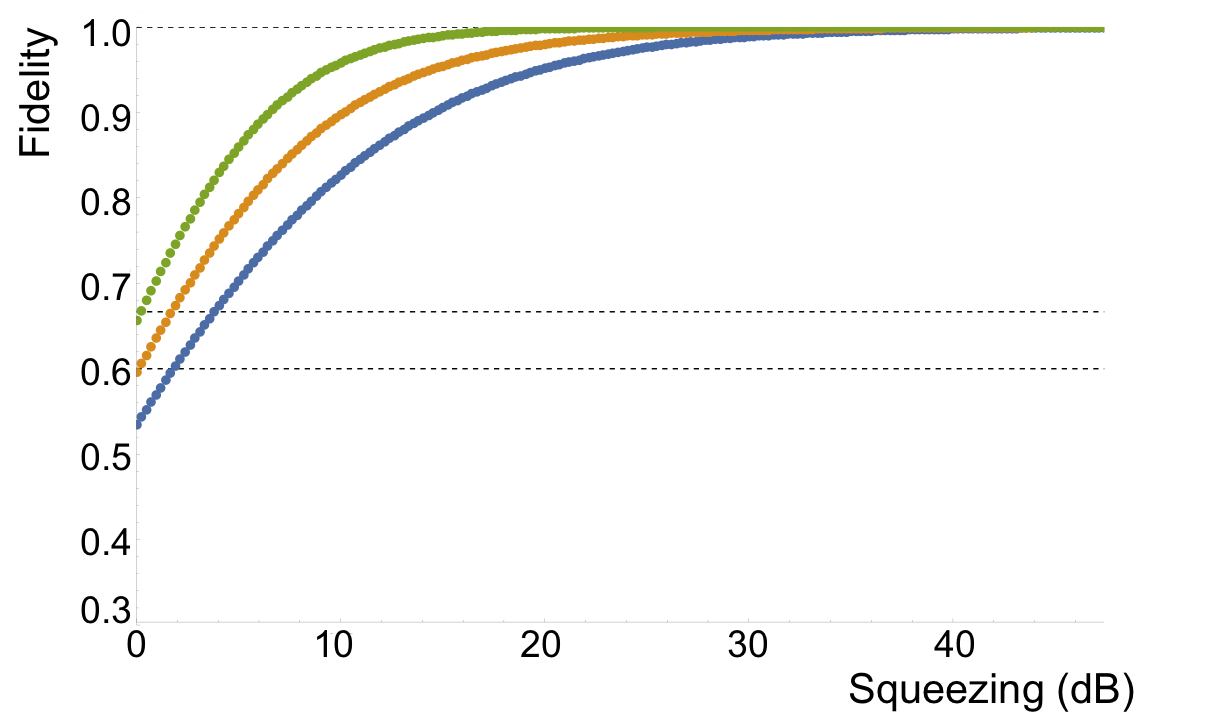}
\caption{\textbf{Theoretical fidelity between the secret and the reconstructed state.} In a 3 access party / 5 player secret sharing protocol, the data were obtained assuming the ratio between the squeezing parameters of the modes used to build the network is fixed, and the overall squeezing level is controlled with a common scaling factor. The horizontal axis is the squeezing level of the most squeezed mode. The top line (green) is the highest fidelity among all the possible access parties while the bottom line (blue) represents the worst. The line in the middle (orange) was obtained by averaging the fidelity over all access parties. }
\label{fig5}
\end{figure}

In order to assess the quality of a secret sharing protocol carried out with our resource, we compute the fidelity between a general input coherent state and the state reconstructed from many measurements of the secret quadratures. We make use of the following formula for the fidelity between two Gaussian states~\cite{Marian2012} \be
 \mathcal{F}=\fr{2}{\sqrt{A+B}-\sqrt{B}}\textrm{exp}\lqu-\alpha^T(V_\textrm{s}+V_\textrm{reS})^{-1}\alpha\rqu,
 \ee where $V_\textrm{s}$ and $V_\textrm{reS}$ are the covariance matrices of the input secret and reconstructed secret, respectively;  $A=\mathrm{det}(V_\textrm{s}+V_\textrm{reS})$, $B=(\mathrm{det}V_\textrm{s}-1)(\mathrm{det}V_\textrm{reS}-1)$;  and $\alpha$ is the difference of the mean amplitudes of the two Gaussian states. When the secret is squeezed vacuum, or when the mean field can be retrieved exactly, $\alpha=0$, which permits the fidelity to be recast as \be
 \mathcal{F}=\fr{2}{\sqrt{A+B}-\sqrt{B}}.
 \label{fide}
 \ee The covariance matrix of the reconstructed secret state and of the initial secret are
 \be
V_\textrm{ReS}= \left(
\begin{array}{cc}
\Delta^2(\hat{x}_{jkl}) & 0 \\
0 & \Delta^2(\hat{p}_{jkl})  \\
\end{array}
\right)\ee and \be V_s= \left(
\begin{array}{cc}
\Delta^2(\hat{x}_s) & 0 \\
0 & \Delta^2(\hat{p}_s)  \\
\end{array}
\right),
\label{noiseS}
\ee
respectively, where $V_\textrm{reS}$ is measured according to equation (\ref{secretsharing}) and $(jkl)$ is any access party. From equation~(\ref{secretsharing}), since the modes are independently squeezed at the beginning, the variances of the reconstructed quadratures are computed as \be\begin{array}{rl} \Delta ^2 \hat{x}_{jkl}=&\Delta ^2 \hat{x}_s + \sum\limits_{i=1} ^5 (a_i ^{jkl}) ^2\Delta ^2 \hat{p}_i ^\mathrm{sqz} \\ \Delta ^2 \hat{p}_{jkl}=&\Delta ^2 \hat{p}_s + \sum\limits_{i=1} ^5 (b_i ^{jkl}) ^2\Delta ^2 \hat{p}_i ^\mathrm{sqz} \end{array}. \label{analyticalFid} \ee Fig.~\ref{fig5} is obtained from equation~(\ref{analyticalFid}) under the assumption that the secret is a coherent state and the squeezing ratio between the modes underlying the network is fixed and follows the distribution seen in Fig.~\ref{figure2}. The overall squeezing is thus adjusted with a common scaling factor. If no squeezing is present in the resource, the best retrieval fidelity among the access parties approaches $2/3$, which is consistent with the teleportation limit achievable with classical resources~\cite{grangier}. Likewise, the average fidelity approaches $3/5$, consistent with the $k/n$ classical limit for threshold schemes of quantum secret sharing~\cite{Tyc}. Both the maximum and the average fidelity, as well as the minimum fidelity across the access parties, approach a value of unity as the overall squeezing level increases. \\

To obtain the black dot-dashed curves in Fig.~\ref{figsecret}, we drew Gaussian-distributed random values with standard deviations matching those of the experimentally measured quadrature squeezing values. Using these random numbers, numerical fidelities are obtained by simulating the secret sharing process with the use of equation~(\ref{fide}).

\section{Acknowledgmenents}

We acknowledge insightful comments and discussion on secret sharing with D. Markham. This work has received funding from the European Union's (EU) Horizon 2020 research and innovation program under Grant Agreement No. 665148, the European Research Council starting grant Frecquam and the French National Research Agency project COMB. C. F. and N. T. are members of the Institut Universitaire de France. J. R. acknowledges support from the European Union through Marie Sklodowska Curie Actions. Y. C. recognizes the China Scholarship Council. X. X. records the National Key Basic Research and Development Program of China under grant 2012CB821302, the National Natural Science Foundation of China under grant 11134003, the National High Technology Research and Development Program of China under grant 2014AA123401.


\bibliographystyle{apsrev}

\end{document}